\documentclass[aps,prl,twocolumn,showpacs,preprintnumbers,amsmath,amssymb,superscriptaddress]{revtex4}%
\usepackage{graphicx}%
\usepackage{dcolumn}
\usepackage{amsmath}
\usepackage{color}
\usepackage{multirow}

\begin{document}


\title{Signature of multi-gap nodeless superconductivity in CaKFe$_4$As$_4$}

\author{P. K. Biswas}
\email[Corresponding author: ]{pabitra.biswas@stfc.ac.uk}
\affiliation{ISIS Pulsed Neutron and Muon Source, STFC Rutherford Appleton Laboratory, Harwell Campus, Didcot, Oxfordshire, OX11 0QX, United Kingdom}

\author{A. Iyo}
\affiliation{National Institute of Advanced Industrial Science and Technology (AIST), 1-1-1 Umezono, Tsukuba, Ibaraki 305-8568, Japan}

\author{Y. Yoshida}
\affiliation{National Institute of Advanced Industrial Science and Technology (AIST), 1-1-1 Umezono, Tsukuba, Ibaraki 305-8568, Japan}

\author{H. Eisaki}
\affiliation{National Institute of Advanced Industrial Science and Technology (AIST), 1-1-1 Umezono, Tsukuba, Ibaraki 305-8568, Japan}

\author{K. Kawashima}
\affiliation{IMRA Material R\&D Co., Ltd., 2-1 Asahi-machi, Kariya, Aichi 448-0032, Japan}

\author{A. D. Hillier}
\affiliation{ISIS Pulsed Neutron and Muon Source, STFC Rutherford Appleton Laboratory, Harwell Campus, Didcot, Oxfordshire, OX11 0QX, United Kingdom}

\begin{abstract}
A newly discovered family of high-$T_{\rm c}$ Fe-based superconductors, $A_e$$A$Fe$_4$As$_4$ ($A_e$ = Ca, Sr, Eu and $A$ = K, Rb, Cs), offers further opportunities to understand unconventional superconductivity in these materials. In this paper, we report on the superconducting and magnetic properties of CaKFe$_4$As$_4$, studied using muon spectroscopy. Zero field muon-spin-relaxation studies carried out on the CaKFe$_4$As$_4$ superconductor do not show any detectable magnetic anomaly at $T_{\rm c}$ or below, implying that time-reversal-symmetry is preserved in the superconducting ground state. The temperature dependence of the superfluid density of CaKFe$_4$As$_4$ is found to be compatible with a two-gap $s$+$s$-wave model with gap values of 8.6(4) and 2.5(3) meV, similar to the other Fe-based superconductors. The presence of two superconducting energy gaps is consistent with theoretical and other experimental studies on this material. The value of the penetration depth at $T=0$~K has been determined as $289(22)$ nm.
\end{abstract}

\maketitle


Recent years have witnessed huge research interest in Fe-based superconductors not only due to higher superconducting transition temperature, $T_{\rm c}$, but also because of the unconventional superconducting properties and interplay with various electronic ground states, such as nematic phase, magnetism \cite{Kamihara,Hsu,Stewart,Chen,Paglione}. This family has a diverse range of structures all containing a common Fe$_2$\textit{An}$_2$ (\textit{An} = P, As, Se, Te) layer, analogous to the CuO$_2$ layers in high-$T_{\rm c}$ cuprates \cite{Bednorz}. $A_e$$A$Fe$_4$As$_4$ ($A_e$ = Ca, Sr, Eu and $A$ = K, Rb, Cs) materials, discovered in 2016, are the new family of Fe-based superconductors with $T_{\rm c}$ of $\approx31-36$ K \cite{Iyo}. Until now, only a few experimental studies have been performed which show that this new family of superconductors is structurally identical to CaFe$_2$As$_2$, with the $A_e$ and $A$ sites forming alternating planes along the crystallographic $c$-axis, separated by FeAs layers \cite {Iyo}. Unlike other Fe-based superconductors, A$_e$AFe$_4$As$_4$ does not show any structural phase transition below room temperature \cite {Meier}. However, Eu-containing A$_e$AFe$_4$As$_4$ show a magnetic transition at $\approx 15$ K \cite{Liu1,Liu2}. Along with high $T_{\rm c}$ values, the A$_e$AFe$_4$As$_4$ family also show a very high upper critical field ($\approx 92$ T for CaKFe$_4$As$_4$) \cite {Kong}, similar to the optimally doped (Ba,K)Fe$_2$As$_2$ \cite{Altarawneh} which is well above the Pauli paramagnetic limit $H_{\rm p}(0)=1.86*T_{\rm c}$ for weak-coupling BCS superconductors. Measurements of the London penetration depth $\lambda(T)$ using a Tunnel-diode resonator (TDR) and tunneling conductance from a Scanning tunneling microscope (STM) spectroscopy show two nodeless gaps in CaKFe$_4$As$_4$ with a larger gap of between 6-9 meV and a smaller gap of between 1-4 meV \cite{Cho}. High-resolution angle-resolved photoemission spectroscopy (ARPES) measurements and density functional theory (DFT) calculations support multi-band superconductivity in CaKFe$_4$As$_4$, in which Cooper pairs form in the electron and the hole bands interacting via a dominant interband repulsive interaction enhanced by Fermi surface (FS) nesting \cite{Mou}. Other thermodynamic and transport measurements on CaKFe$_4$As$_4$, such as resistivity, thermoelectric power, Hall effect, magnetization, and specific heat, etc. all show similar properties to the optimally doped (Ba$_{1-x}$K$_x$)Fe$_2$As$_2$ \cite{Meier}. This fully ordered and stoichiometric A$_e$AFe$_4$As$_4$ family of superconductors with high-$T_{\rm c}$ values therefore provide new and unique opportunities to test and compare the experimental and theoretical understanding of the Fe-based superconductors. In addition, the ordered nature of the structure makes theoretical models easier to develop. In this respect, it is of critical importance to understand the superconducting and magnetic properties of CaKFe$_4$As$_4$ and obtain detailed knowledge of the superconducting order parameters.

In this article, we report the results from muon spin rotation and relaxation ($\mu$SR) studies on the superconducting and magnetic properties of CaKFe$_4$As$_4$ at the microscopic level. We have investigated the symmetry of the superconducting gap of CaKFe$_4$As$_4$ by measuring the magnetic penetration depth as a function of temperature $\lambda(T)$. Our results show that CaKFe$_4$As$_4$ is a nodeless multi-gap superconductor with gap sizes of 8.6(4) and 2.5(3) meV. We also obtained the penetration depth at 0 K, $\lambda(0)=289(22)$~nm. Using ZF-$\mu$SR technique we confirm that CaKFe$_4$As$_4$ is paramagnetic in the normal state. It further shows that the superconducting state of CaKFe$_4$As$_4$ preserves time-reversal-symmetry (TRS).


Polycrystalline samples of CaKFe$_4$As$_4$ were synthesized using the stainless steel (SUS) pipe and cap method described in ref.~\onlinecite{Iyo}. Sample characterization measurements (not shown here) were performed at the National Institute of Advanced Industrial Science and Technology (AIST), Japan. Zero-field (ZF)- and transverse-field (TF)-$\mu$SR experiments were carried out using the MuSR spectrometer \cite{isis} at the ISIS facility, STFC Rutherford Appleton Laboratory, UK. Powder samples of CaKFe$_4$As$_4$ were pressed into 6 pellets of diameter 6 mm and thickness $\approx$1 mm each. For measurements in ZF-mode, a mosaic of the 6 pellets were mounted onto a 99.995\% pure silver plate. Any muon stopping in the silver plate gave a constant background signal over the temperature range of interest. For the measurements in TF-mode, hematite slabs were positioned immediately behind the sample. Muons stopped in the hematite slabs depolarized immediately and hence the background in the data collected was significantly reduced. In the TF-$\mu$SR experiments, 40 mT field was applied (above $T_{\rm c}$) perpendicular to the initial muon spin polarization. The samples were then cooled to base temperature (1.5 K) in a magnetic field and the $\mu$SR spectra were collected upon warming the sample. The typical counting statistics were $\sim30$ million muon decays per data point. The ZF- and TF-$\mu$SR data were analyzed using the free software packages Mantid~\cite{mantid} and MUSRFIT \cite{Suter}.


\begin{figure}[htb]
\includegraphics[width=1.0\linewidth]{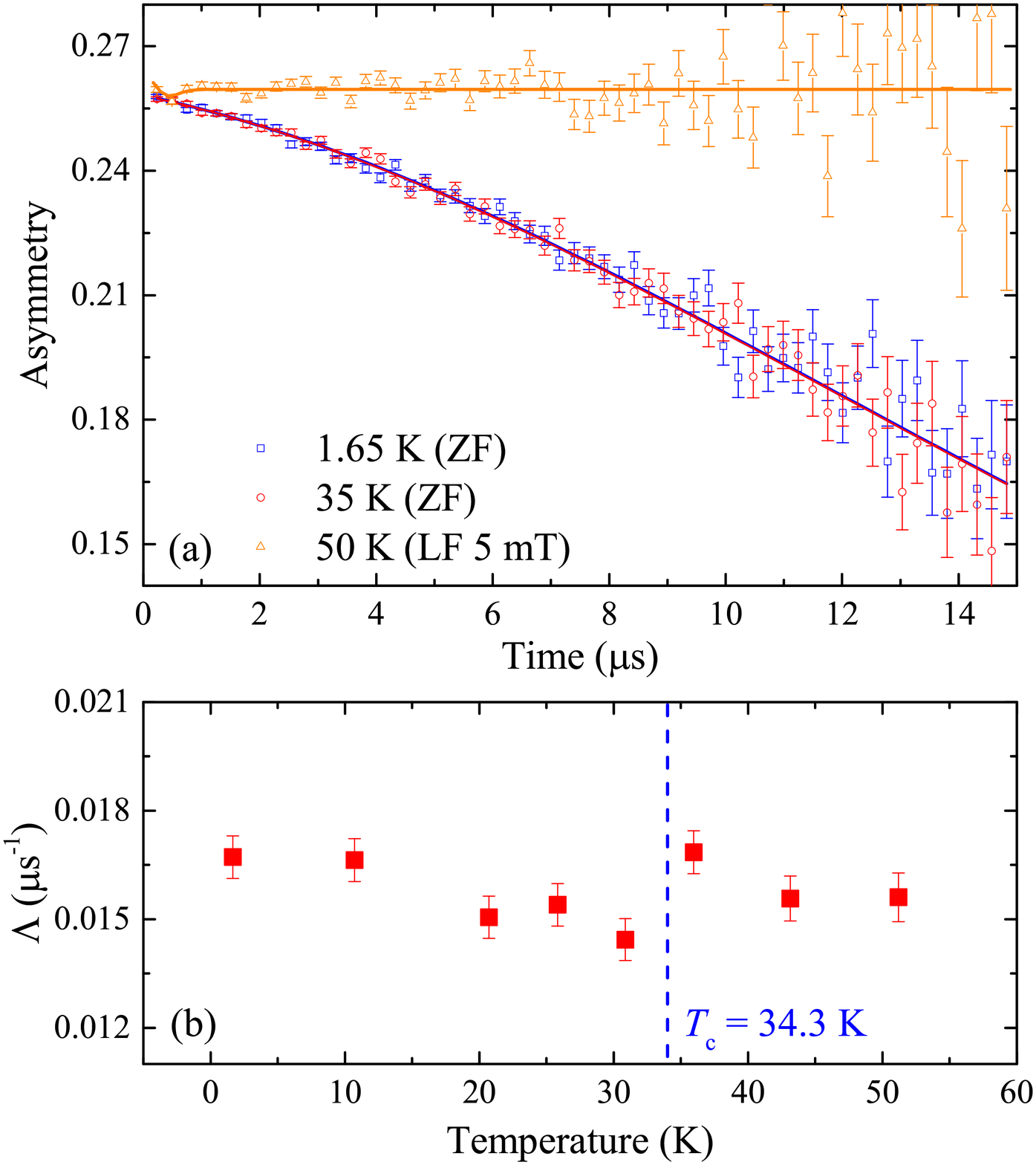}
\caption{(Color online) (a) Zero-field ($T=1.65$ and $T=35$ K) and longitudinal-field ($T=50$ K and $\mu_0H=5$ mT) $\mu$SR time spectra for CaKFe$_4$As$_4$. The solid lines are the fits to the data using Eq.~\ref{eq:KT_ZFequation}, described in the text. (b) Temperature dependence of the muon spin relaxation rate $\Lambda$, extracted from the ZF-$\mu$SR time spectra.}
 \label{fig:ZF}
\end{figure}

ZF-$\mu$SR measurements were performed to investigate the magnetic properties of CaKFe$_4$As$_4$ superconductor, especially to look for any additional magnetic signal arising spontaneously with the onset of superconductivity, associated with time-reversal-symmetry breaking fields. Figure \ref{fig:ZF} (a) shows the ZF-$\mu$SR time spectra measured at $T=1.65$ and 35~K. ZF data collected above and below $T_{\rm c}$ show negligible difference, implies that there is no additional relaxation of the $\mu$SR signal appearing in the superconducting state of CaKFe$_4$As$_4$. The data were analyzed using the Kubo-Toyabe relaxation function~\cite{Kubo} multiplied by an exponential decay function with a decay rate $\Lambda$,

\begin{multline}
A(t)= A(0)\left\{\frac{1}{3}+\frac{2}{3}\left(1-a^2t^2\right){\exp}\left(-\frac{a^2t^2}{2}\right)\right\} \\
{\exp}(-\Lambda t),
 \label{eq:KT_ZFequation}
\end{multline}
where $A(0)$ is the initial asymmetry, and $a$ and $\Lambda$ are muon spin relaxation rates of the nuclear and electronic moments, respectively. Since we expect that the contribution from the nuclear moments will be similar above and below $T_{\rm c}$, $a$ was kept as a common parameter for all temperature data. The fits yield $a = 0.045(2)$~$\mu$s$^{-1}$ which is very small and reflects the presence of random local fields arising from the nuclear moments within CaKFe$_4$As$_4$. Figure \ref{fig:ZF} (b) show that $\Lambda$ is nearly temperature independent across $T_{\rm c}$ and therefore rules out the presence of any magnetic anomaly in the superconducting state of CaKFe$_4$As$_4$. The small values of $\Lambda$ are consistent with the presence of diluted and randomly oriented electronic moments in this material. A 5 mT longitudinal field (LF) is sufficient to decouple the remaining moments completely (see Fig.\ref{fig:ZF} (a)), which further suggests that these moments are static in nature. All the ZF-$\mu$SR time spectra could be fitted with a single asymmetry function, implying that CaKFe$_4$As$_4$ is fully paramagnetic in the normal state.

\begin{figure}[htb]
\includegraphics[width=1.0\linewidth]{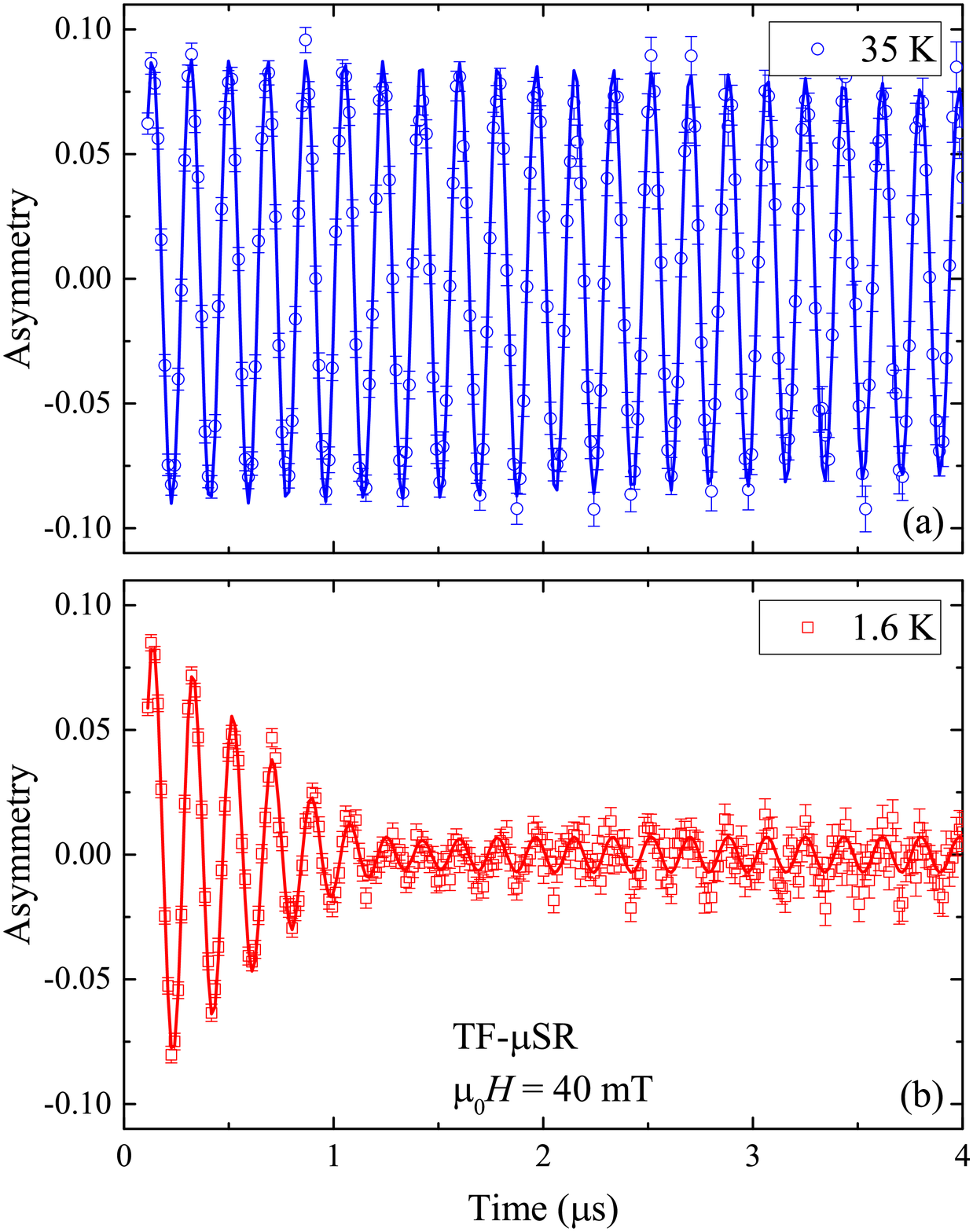}
\caption{(Color online) TF-$\mu$SR time spectra of CaKFe$_4$As$_4$, collected at 1.6 and 35~K in a transverse field of 40 mT. The solid lines are the fits to the data using the Eq.~\ref{Depolarization_Fit}.}
 \label{fig:Asy_TF}
\end{figure}

Figure~\ref{fig:Asy_TF} shows the TF-$\mu$SR time spectra collected at 1.6 and 35~K in a TF of 40 mT. As expected, muon spin depolarisation rate in the superconducting state (1.6 K data) is much stronger than in the normal state (35 K data) due to the spatial field modulation of the flux line lattice, formed in the mixed state of CaKFe$_4$As$_4$. All these responses can be described by an oscillatory term with a Gaussian envelope,
\begin{multline}
\label{Depolarization_Fit}
A^{TF}(t)=A(0)\exp\left(-\sigma^{2}t^{2}\right/2)\cos\left(\gamma_\mu \left\langle B\right\rangle t +\phi\right) \\
+A_{\rm bg}(0)\cos\left(\gamma_\mu B_{\rm bg}t +\phi\right),
\end{multline}
where $A(0)$ and $A_{\rm bg}$(0) are the initial asymmetries of the sample and background signals, $\gamma_{\mu}/2\pi=135.5$~MHz/T is the muon gyromagnetic ratio~\cite{Sonier}, $\left\langle B\right\rangle$ and $B_{\rm bg}$ are the internal and background magnetic fields, $\phi$ is the initial phase of the muon precession signal, and $\sigma$ is the Gaussian muon spin relaxation rate. $\sigma$ represents the second moment of the internal field distribution. It is worthwhile to mention here that our background signals are very small due to the hematite slabs mounted immediately after the sample. Any muons missing the sample implanted into the hematite slabs which then depolarize quickly within few nanoseconds and appear as an asymmetry loss in the TF-$\mu$SR signals. A very small background signal most probably originates from a small number of muons hitting the Ag sample holder and is assumed to be non-relaxing over the muon time window, an assumption supported by ZF-$\mu$SR.

\begin{figure}[htb]
\includegraphics[width=1.0\linewidth]{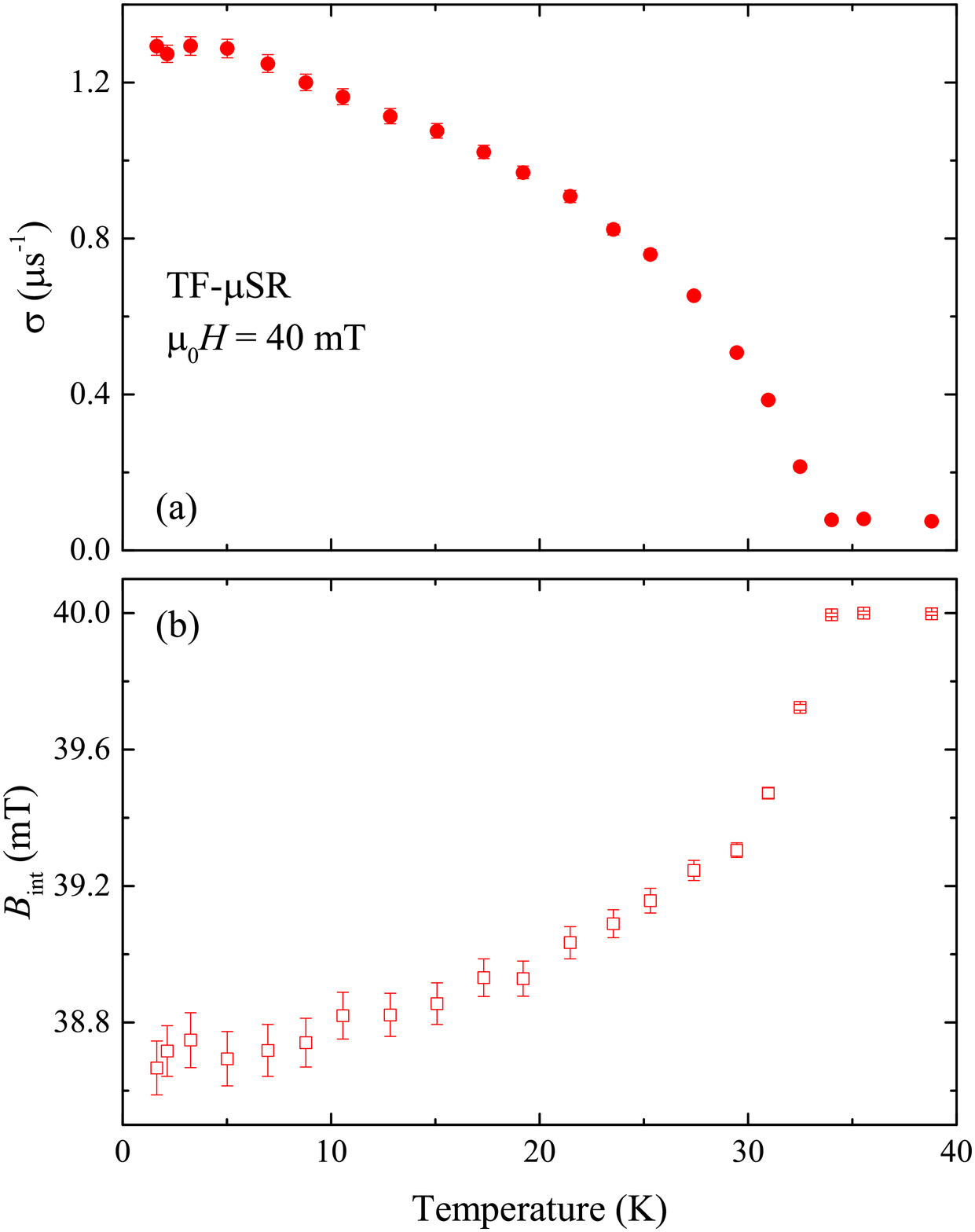}
\caption{(Color online) (a) Temperature dependence of the muon spin relaxation rate $\sigma$ of CaKFe$_4$As$_4$. (b) Temperature dependence of the internal field strength of CaKFe$_4$As$_4$ for an applied field of 40 mT.}
 \label{fig:sigma_field}
\end{figure}

Figure~\ref{fig:sigma_field} (a) shows the temperature dependence of $\sigma$ for CaKFe$_4$As$_4$ which exhibits a pronounced increase below $T_{\rm c}$. Figure~\ref{fig:sigma_field} (b) shows the expected diamagnetic shift below $T_{\rm c}$ in the form of internal magnetic field strength at the muon site in CaKFe$_4$As$_4$. $\sigma$ can be expressed as $\sigma=\left(\sigma^{2}_{\rm sc} + \sigma^{2}_{\rm nm}\right)^{\frac{1}{2}}$, where $\sigma_{sc}$ is the superconducting contribution to the relaxation rate due to the inhomogeneous field variation across the flux line lattice and $\sigma _{nm}$ is the nuclear magnetic dipolar contribution which is assumed to be temperature independent.

\begin{figure}[htb]
\includegraphics[width=1.0\linewidth]{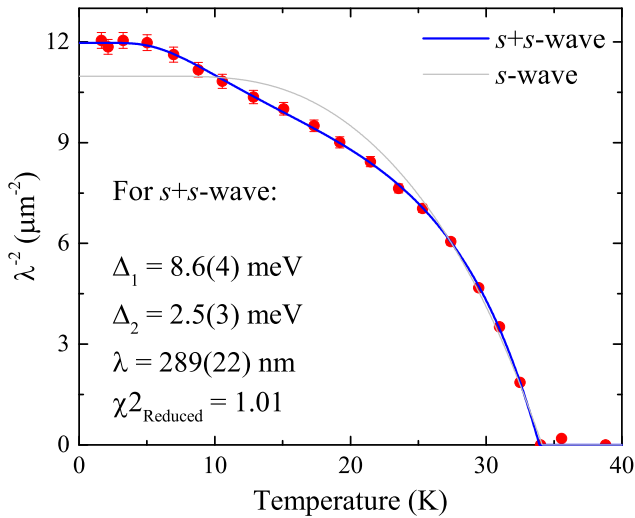}
\caption{(Color online) Temperature dependence of $\lambda^{-2}$ for CaKFe$_4$As$_4$. The solid curves are the fits to the data using a single-gap \textit{s}-wave and two-gap $s$+$s$-wave models.}
 \label{fig:lambda}
\end{figure}

In light of Ginzburg-Landau treatment of the vortex state, Brandt \cite{Brandt2} has shown that for a superconductor with a hexagonal vortex lattice and very high upper critical field, $\sigma_{\rm sc}$ is related to the penetration depth, $\lambda$ by the simplified Brandt equation,

\begin{equation}
\frac{\sigma_{sc}\left(T\right)}{\gamma_\mu}=0.06091\frac{\Phi_0}{\lambda^{2}\left(T\right)},
\end{equation}
where $\Phi_0=2.068\times10^{-15}$~Wb is the flux quantum~\cite{Brandt2}. $\lambda^{-2}(T)$ is proportional to the effective superfluid density, $\rho_s\propto\lambda^{-2}$ and hence carries the information about the symmetry and size of a superconducting gap. Figure~\ref{fig:lambda} shows the temperature dependence of $\lambda^{-2}$ for CaKFe$_4$As$_4$. The fits to the $\lambda^{-2}(T)$ data were made using either a single gap or a combination of two gaps models using the following functional form~\cite{Carrington, Padamsee}:

\begin{equation}
\label{two_gap}
\frac{\lambda^{-2}\left(T\right)}{\lambda^{-2}\left(0\right)}=\omega\frac{\lambda^{-2}\left(T, \Delta_{1}(0)\right)}{\lambda^{-2}\left(0,\Delta_{1}(0)\right)}+(1-\omega)\frac{\lambda^{-2}\left(T, \Delta_{2}(0)\right)}{\lambda^{-2}\left(0,\Delta_{2}(0)\right)},
\end{equation}
where $\lambda\left(0\right)$ is the value of the penetration depth at $T=0$~K, $\Delta_{\rm i}(0)$ is the value of the $i$-th ($i=1$ or 2) superconducting gap at $T=0$~K and $\omega$ is the weighting factor of the first gap.

Each component of Eq.~\ref{two_gap} can be expressed within the local London approximation ($\lambda \gg \xi$)~\cite{Tinkham,Prozorov} as

\begin{equation}
\frac{\lambda^{-2}\left(T, \Delta_{0,i}\right)}{\lambda^{-2}\left(0, \Delta_{0,i}\right)}=1+\frac{1}{\pi}\int^{2\pi}_{0}\int^{\infty}_{\Delta_{\left(T,\varphi\right)}}\left(\frac{\partial f}{\partial E}\right)\frac{ EdE d\varphi}{\sqrt{E^2-\Delta_i\left(T,\varphi\right)^2}},
\end{equation}
where $f=\left[1+\exp\left(E/k_{\rm B}T\right)\right]^{-1}$ is the Fermi function, $\varphi$ is the angle along the Fermi surface, and $\Delta_i\left(T,\varphi\right)=\Delta_{0, i}\delta\left(T/T_c\right)g\left(\varphi\right)$, where $g\left(\varphi\right)$ describes the angular dependence of the gap. The temperature dependence of the gap can be approximated by the expression $\delta\left(T/T_{\rm c}\right)=\tanh\left\{1.82\left[1.018\left(T_{\rm c}/T-1\right)\right]^{0.51}\right\}$~\cite{Carrington}.

The solid curves shown in Fig.~\ref{fig:lambda} are fits to the $\lambda^{-2}(T)$ data from CaKFe$_4$As$_4$ using a single gap $s$-wave and two-gap $s+s$-wave models. For simplicity, we have not shown the fitted curves corresponding to anisotropic $s$- and $s+d$-wave models in Fig.~\ref{fig:lambda}. All the fitted parameters are summarized in Table~\ref{table_of_gapratios}. The two-gap \textit{s}+\textit{s}-wave gap model gives a lower $\chi_{\rm reduced}^2$ value than any other above mentioned modelsmodels, and gives a much better fit to the data compared to the single-gap \textit{s}-wave model with gap values of 8.6(4) and 2.5(3) meV and $\lambda(0)=289(22)$ nm. Therefore it can be concluded that CaKFe$_4$As$_4$ is a multiple gap superconductor which is consistent with other layered Fe-based superconductors~\cite{Khasanov,Biswas2,Biswas3,Paglione}. Some other well-known examples of multi-gap superconductivity are represented by MgB$_2$, NbSe$_2$, and Lu$_2$Fe$_3$Si$_5$~\cite{Choi,Yokoya,Nakajima,Biswas4}. The two-gap model, used here to analyze our final results, is a very simple approach to describe multiple superconducting gaps in the pairing symmetry data and has been used extensively in the past. For example, a similar model has also been used by Ch. Niedermayer \textit{et al} to investigate the symmetry and values of the superconducting gaps in MgB$_2$ \cite{Niedermayer}.

\begin{table}
\caption{Fitted parameters to the $\lambda^{-2}(T)$ data of CaKFe$_4$As$_4$ using the different models as described in the text.}
\label{table_of_gapratios}
\begin{center}
\begin{tabular}[t]{llll}\hline\hline
Model&$g\left(\varphi\right)$& Gap value (meV)&$\chi ^2$\\\hline
$s$-wave&1&$\Delta$=6.9(1)&12.6\\
$s+s$-wave&1&$\Delta_1$=8.6(4), $\Delta_2$=2.5(3) and $\omega=0.68(4)$&1.05\\
anisotropic $s$-wave&$\left(s+\cos4\varphi\right)$&$\Delta=7.0(1)$ with $s=0.71(3)$&1.4\\
$s+d$-wave&1, $\left|\cos\left(2\varphi\right)\right|$&$\Delta_1$=10(1), $\Delta_2$=6(1) and $\omega=0.44(8)$&1.56\\\hline\hline
\end{tabular}
\par\medskip\footnotesize
\end{center}
\end{table}

The superconducting gap values obtained from the two-gap model fits are in excellent agreement with the reported values obtained from the STM measurements for this material \cite{Cho} and also consistent with other layered Fe-based superconductors, especially from the $\mu$SR and ARPES measurements of optimally doped (Ba,K,Rb)Fe$_2$As$_2$ system \cite{Khasanov1,Guguchia}. These are, however, more than double the values reported from the TDR measurements of CaKFe$_4$As$_4$ \cite{Cho}. Our finding of two superconducting gaps in CaKFe$_4$As$_4$ is not unsurprising and is very much consistent with other members of Fe-based superconductors. It is now an universal phenomena across all Fe-based superconductors, apart from a small number of exceptions, such as the highly electron-doped regime of (Ba,K,Rb)Fe$_2$As$_2$ system.


In summary, ZF- and TF-$\mu$SR measurements have been performed on CaKFe$_4$As$_4$, a member of the newly discovered family of high-$T_{\rm c}$ Fe-based superconductors $A_e$$A$Fe$_4$As$_4$ ($A_e$ = Ca, Sr, Eu and $A$ = K, Rb, Cs). ZF-$\mu$SR data confirm that there is no magnetic anomaly in the superconducting ground state of CaKFe$_4$As$_4$. The temperature dependence of the magnetic penetration depth is found to be strongly compatible with a two-gap $s$+$s$-wave model with gap values, $\Delta_{1}(0)=8.6(4)$, $\Delta_{2}(0)=2.5(3)$~meV with weighting factor $\omega=0.68(4)$ to the large gap. We obtain the penetration depth, $\lambda(0)=289(22)$~nm. Our results show that CaKFe$_4$As$_4$ is a fully developed nodeless multi-gap superconductor and all the fitted gap parameters are very similar to the optimally doped (Ba,K,Rb)Fe$_2$As$_2$ systems. Further studies are in progress to explore the pressure evolution of the symmetry of the superconducting gaps in this materials.

The $\mu$SR experiments were performed at the ISIS Pulsed Neutron and Muon Source, STFC Rutherford Appleton Laboratory, (STFC, United Kingdom).


\begin{thebibliography}{99}
%
\bibitem{Kamihara} Y.~Kamihara, T.~Watanabe, M.~Hirano, and H.~Hosono, J.~Am.~Chem.~Soc. {\bf 130}, 3296 (2008).
%
\bibitem{Hsu} F.-C.~Hsu, J.-Y.~Luo, K.-W.~Yeh, T.-K.~Chen, T.-W.~Huang, P.~M.~Wu, Y.-C.~Lee, Y.-L.~Huang, Y.-Y.~Chu, D.-C.~Yan, and M.-K.~Wu, Proc.~Natl.~Acad.~Sci.~USA {\bf 105}, 14262 (2008).
%
\bibitem{Stewart} G. R. Stewart, Rev. Mod. Phys. {\bf 83}, 1589 (2011).
%
\bibitem{Chen} X. Chen, P. Dai, D. Feng, T. Xiang, F. C. Zhang, Natl. Sci. Rev. \textbf{1}, 371 (2014).
%
\bibitem{Paglione} J. Paglione and R. L. Greene, Nat. Phys. \textbf{6}, 645 (2010).
%
\bibitem{Bednorz} J.~G.~Bednorz, and K.~A.~Muller, Z.~Physik~B {\bf 64}, 189 (1986).
%
\bibitem{Iyo} A. Iyo, K. Kawashima, T. Kinjo, T. Nishio, S. Ishida, H. Fujihisa, Y. Gotoh, K. Kihou, H. Eisaki, and Y. Yoshida, J. Am. Chem. Soc. \textbf{138}, 3410 (2016).
%
\bibitem{Meier} W. R. Meier, T. Kong, U. S. Kaluarachchi, V. Taufour, N. H. Jo, G. Drachuck, A. E. B\"ohmer, S. M. Saunders, A. Sapkota, A. Kreyssig, M. A. Tanatar, R. Prozorov, A. I. Goldman, S. L. Bud'ko, and P. C. Canfield, arXiv:1605.05617 (2016).
%
\bibitem{Liu1} Y. Liu, Y.-B. Liu, Z.-T. Tang, H. Jiang, Z.-C. Wang, A. Ablimit, W.-H. Jiao, Q. Tao, C.-M. Feng, Z.-A. Xu, and G.-H. Cao, arxiv:1605.04396 (2016).
%
\bibitem{Liu2} Y. Liu, Y.-B. Liu, Q. Chen, Z.-T. Tang, W.-H. Jiao, Q. Tao, Z.-A. Xu, and G.-H. Cao, arxiv:1605.09007 (2016).
%
\bibitem{Kong} T. Kong, F. F. Balakirev, W. R. Meier, S. L. Bu\'dko, A. Gurevich, and P. C. Canfield, arXiv:1606.02241 (2016).
%
\bibitem{Altarawneh} M. M. Altarawneh, K. Collar, C. H. Mielke, N. Ni, S. L. Bu\'dko, and P. C. Canfield, Phys. Rev. B \textbf{78}, 220505 (2008).
%
\bibitem{Cho} K. Cho, A. Fente, S. Teknowijoyo, M. A. Tanatar, T. Kong, W. Meier, U. Kaluarachchi, I. Guillam\'on, H. Suderow, S. L. Bu\'dko, P. C. Canfield, and R. Prozorov, arXiv:1606.06245 (2016).
%
\bibitem{Mou} D. Mou, T. Kong, W. R. Meier, F. Lochner, L.-L. Wang, Q. Lin, Y. Wu, S. L. Bu\'dko, I. Eremin, D. D. Johnson, P. C. Canfield, A. Kaminski, arXiv:1606.05643 (2016).
%
\bibitem{isis} www.isis.stfc.ac.uk/instruments/musr/.
%
%
\bibitem{mantid} www.mantidproject.org.
%
\bibitem{Suter} A.~Suter, and B.~M.~Wojek, Physics~Procedia {\bf 30}, 69 (2012).
%
\bibitem{Kubo} R.~Kubo, Hyperfine~Interact. {\bf 8}, 731 (1981).
%
\bibitem{Sonier} J.~E.~Sonier, J.~H.~Brewer, and R.~F.~Kiefl, Rev.~Mod.~Phys. {\bf 72}, 769 (2000).
%
%
\bibitem{Brandt2} E.~H.~Brandt, Phys.~Rev.~B {\bf 68}, 054506 (2003).
%
\bibitem{Carrington} A.~Carrington, and F.~Manzano, Physica~C {\bf 385}, 205 (2003).
%
\bibitem{Padamsee} H.~Padamsee, and J.~E.~Neighbor, and C.~A.~Shiffman, J.~Low~Temp.~Phys. {\bf 12}, 387 (1973).
%
\bibitem{Tinkham} M.~Tinkham, Introduction~to~Superconductivity (McGraw-Hill, New~York, 1975).
%
\bibitem{Prozorov} R.~Prozorov, and R.~W.~Giannetta, Supercond.~Sci.~Technol. {\bf 19}, R41 (2006).
%
\bibitem{Khasanov} R.~Khasanov, K.~Conder, E.~Pomjakushina, A.~Amato, C.~Baines, Z.~Bukowski, J.~Karpinski, S.~Katrych, H.-H.~Klauss, H.~Luetkens, A.~Shengelaya, N.~D.~Zhigadlo, Phys.~Rev.~B {\bf 78}, 220510(R) (2008).
%
\bibitem{Biswas2} P.~K.~Biswas, G.~Balakrishnan, D.~McK.~Paul, C.~V.~Tomy, M.~R.~Lees, A.~D.~Hillier, Phys.~Rev.~B {\bf 81}, 092510 (2010).
%
\bibitem{Biswas3} P.~K.~Biswas, A.~Krzton-Maziopa, R.~Khasanov, H.~Luetkens, E.~Pomjakushina, K.~Conder, and A.~Amato, Phys.~Rev.~Lett. {\bf 110}, 137003 (2013).
%
\bibitem{Niedermayer} Ch. Niedermayer, C. Bernhard, T. Holden, R. K. Kremer, and K. Ahn, Phys.~Rev.~B {\bf 65}, 094512 (2002).
%
\bibitem{Choi} H.~J.~Choi, D.~Roundy, H.~Sun, M.~L.~Cohen, and S.~G.~Louie, Nature~(London) {\bf 418}, 758 (2002), and references therein.
%
\bibitem{Yokoya} T.~Yokoya, T.~Kiss, A.~Chainani, S.~Shin, M.~Nohara, H.~Takagi, Science {\bf 294}, 2518 (2001).
%
\bibitem{Nakajima} Y.~Nakajima, T.~Nakagawa, T.~Tamegai, and H.~Harima, Phys.~Rev.~Lett. {\bf 100}, 157001 (2008).
%
\bibitem{Biswas4} P.~K.~Biswas, G.~Balakrishnan, D.~McK. Paul, M.~R. Lees, and A.~D. Hillier, Phys.~Rev.~B {\bf 83}, 054517 (2011).
%
\bibitem{Khasanov1} R.~Khasanov, D. V. Evtushinsky, A. Amato, H.-H. Klauss, H. Luetkens, Ch. Niedermayer, B. B\"uchner, G. L. Sun, C. T. Lin, J. T. Park, D. S. Inosov, and V. Hinkov, Phys.~Rev.~Lett. {\bf 102}, 187005 (2009).
%
\bibitem{Guguchia} Z. Guguchia, A. Amato, J. Kang, H. Luetkens, P.K. Biswas, G. Prando, F. von Rohr, Z. Bukowski, A. Shengelaya, H. Keller, E. Morenzoni, Rafael M. Fernandes and R. Khasanov, Nat. Commun. \textbf{6} 8863 (2015).





\end{thebibliography}
\end{document}